\documentclass[pra,twocolumn,showpacs,preprintnumbers,superscriptaddress]
{revtex4-2}
\usepackage{times}
\usepackage{bm}
\usepackage{graphicx}
\usepackage{amsbsy}
\usepackage{amssymb}
\usepackage{amsmath}
\usepackage{amsfonts}
\usepackage{amsthm}
\usepackage{tikz}
\usepackage{caption}
\usepackage{refstyle}
\usepackage{braket}
\usepackage{comment}
\usepackage[colorlinks=true, citecolor=blue, linkcolor=blue, urlcolor=blue]{hyperref}
\usetikzlibrary{shapes, arrows, positioning, arrows, automata, quotes}
\begin{document}
\theoremstyle{plain}
\newtheorem{theorem}{Theorem}
\newtheorem{lemma}[theorem]{Lemma}
\newtheorem{corollary}[theorem]{Corollary}
\newtheorem{proposition}[theorem]{Proposition}
\newtheorem{conjecture}[theorem]{Conjecture}
\theoremstyle{definition}
\newtheorem{definition}[theorem]{Definition}
\theoremstyle{remark}
\newtheorem*{remark}{Remark}
\newtheorem{example}{Example}
\counterwithin*{example}{section}
\title{Construction of PPT entangled state and its detection by using second-order moment of the partial transposition}
\author{Rohit Kumar, Satyabrata Adhikari}
\email{rohitkumar@dtu.ac.in, satyabrata@dtu.ac.in} \affiliation{Department of Applied Mathematics, Delhi Technological
University, Delhi, Delhi, India}
\begin{abstract}
We adopt a formalism by which we construct and detect a new family of positive partial transpose entangled states in $d_1\otimes d_2$ dimensional system. Our detection method is based on the second order moment $p_2(\rho^{T_B})$ as it is very easy to calculate and may be realizable in laboratory. We show that if the second order moment $p_2(\rho^{T_B})$ in $d_1\otimes d_2$ dimensional system satisfy $p_2(\rho^{T_B})\leq\frac{1}{d_1 d_2-1}$, then the state is a PPT state. We also derive an equivalent condition on the bloch vector. Then, we construct a quantum state by considering the mixture of a separable and an entangled state and obtain a condition on the mixing parameter for which the mixture represents a PPTES. Finally, applying our results, we have shown that the distillable key rate of the private state, prepared through our prescription, is positive. It suggests that our result also has potential applications in quantum cryptography.
\end{abstract}
 \maketitle
\section{Introduction}
Quantum entanglement stands as a cornerstone of quantum mechanics, underpinning numerous applications in quantum information science, including quantum computing, quantum teleportation, and quantum cryptography. The ability to detect and characterise entanglement in quantum states is crucial \cite{Horodecki2009}. Yet, it remains a challenging problem for both multipartite systems and bipartite higher-dimensional systems in which PPT entangled states (entangled states that yield no distillable entanglement under local operations and classical communication (LOCC)) exist \cite{P_Horodecki1997}. PPT entangled states are also known as bound entangled states. The construction of PPT entangled states \cite{bruss2000,zhao2016,halder2019} and their detection are one of the important problems in quantum information theory. Traditional criteria, such as the Positive Partial Transpose (PPT) criterion introduced by Peres and the Horodecki's \cite{peres, Horodeckis1996}, provide powerful tools for identifying separability in low-dimensional systems, but they are insufficient for detecting all forms of entanglement, especially PPT entangled states in higher dimensions. PPT criterion is also known as Peres-Horodecki criterion and may also be stated as a necessary and sufficient criterion to detect entangled states in $2\otimes 2$ and $ 2\otimes 3$ dimensional systems. For higher dimensional systems, it provides only necessary condition for separability but not the sufficient condition. PPT criterion is based on the partial transposition map that act on the quantum state described by the density matrices. After the application of the partial transposition operation on the density matrices, if the eigenvalues of the partial transposed matrices at the output, have at least one negative eigenvalues then the corresponding bipartite quantum states are entangled. It may be noted that the partial transposition map is not a 2-positive map and thus it does not represent a completely positive map. Hence it may not be implementable in an experiment. But it clears a fact that if one identify the positive but not completely positive maps then they can take part in the detection of entanglement \cite{rohira2021,scala2024}. Realignment criterion \cite{rudolph2003,chen2002} is also an important entanglement detection criterion, which may detect PPT entangled state in a more efficient manner than the Peres-Horodecki criterion. The problem with the realignment criterion is that it works in a nice way, but theoretically and may not be possible to implement it in an experiment. Another limitation of Realignment criterion is that it provides only a sufficient condition for entanglement, but not a necessary one. Range criterion \cite{P_Horodecki1997} is another entanglement detection criterion. This criterion can detect some bound entangled states that Peres-Horodecki criterion fails to detect. However, like Peres-Horodecki PPT criterion, for higher dimensional systems (higher than $2\otimes 2$ and $2\otimes 3$ dimensional systems), it provides only the sufficient condition for separability but not the necessary one. Apart from these limitations, the Range criterion is dificult to apply, especially for higher dimensional systems as it requires spanning set of product states, which is not algorithmically efficient for large systems. Another important entanglement detection criterion is Reduction criterion \cite{horodecki3}. This criterion provides a necessary condition for separability of a bipartite quantum state. So, if a quantum state violate this condition, the state is an entangled state. The  Reduction criterion is implied by Peres-Horodecki PPT criterion i.e. every PPT state automatically satisfies the Reduction criterion. Hence, it is weaker than the PPT criterion. Moreover, some entangled states also satisfies the reduction inequalities, therefore, it provides only a necessary but not sufficient condition for separability. Similar to the Peres-Horodecki criterion it also fails to detect PPT entangled (bound entangled) states. The Majorization criterion \cite{nielsen2001} is another important separability criterion. Similar to the Reduction criterion it provides only a necessary but not sufficient condition for separability. This criterion requires exact eigenvalue spectra of the composite and the reduced density matrices, which makes it difficult to apply, especially for higher-dimensional systems. The Majorization criterion is stronger than the Reduction criterion, yet weaker than the Peres-Horodecki criterion. Entanglement witnesses, on the other hand, offer a complementary approach by providing observables that can certify entanglement through negative expectation values, though constructing optimal witnesses remains nontrivial \cite{Horodeckis1996,Terhal2000}. Several of the above criteria are discussed in \cite{Bengtsson2006} and examined in \cite{sauer2022}. In \cite{sauer2022} the authors characterise the performance or effectiveness by numerically investigating Euclidean volume ratios between non-violating states and full state space in various bipartite quantum systems.\\
State tomography is a good method to gain knowledge about the system, but its drawback is that for higher-dimensional systems, we need to perform a large number of measurements. Thus, one can detect an entangled state in higher higher-dimensional system using the state tomography method, but at the price of an unlimited number of measurements. We can overcome this problem if we use the partial information of the system to detect an entangled state, and this idea motivated us to construct a witness operator that may be implemented in an experiment very easily \cite{goswami2019}. There exists another method known as the method of moments that may be useful in the detection of entangled states. The advantage of this method is that it can be estimated using shadow tomography in a more efficient way than quantum state tomography. Elben et al. \cite{elben2020} proposed a moment-based method to detect bipartite entanglement. They have used the moments of the partial transposition of the density matrix. Neven et al. \cite{neven2021} proposed an ordered set of experimentally accessible conditions for detecting entanglement in mixed states. The above-mentioned works can only detect negatively partial transposed entangled states. Recently, one of the authors of this work has studied the entanglement detection problem and found a way to detect both negative partial transpose entangled states and PPT entangled states through partial realigned moments \cite{aggarwal2024}. But in recent times, it looks like partial transposition moments are more experimentally friendly than the partial realigned moment, so we take up this challenge to detect PPT entangled states using partial transposition moments.\\
In this work, we use the second-order moment $(p_2)$ of the partial transpose of a bipartite state to detect positive partial transpose states. We first establish a sufficient condition involving $p_2$ and the system's dimension that guarantees a state to be a PPT state. Our contributions then build upon established results, such as the inequality $ p_2^2 \leq p_3 $ for PPT states \cite{elben2020}, to derive novel bounds and the violation of those bounds may help in the detection of entangled states. We further strengthen the framework by deriving a lower bound of $p_2$ for arbitrary bipartite PPT states, complementing the existing upper bound and offering a more complete characterisation of the partial transpose's spectral properties.\\
Building on these foundations, we consider convex mixtures of separable states and PPT entangled states detectable by a given witness operator $W$. For such mixtures $ \rho = p \rho_{SEP} +(1-p) \rho_{PPTES} $, where $\rho_{SEP} $ represents separable and $ \rho_{PPTES} $ represents PPT entangled states, we derive an explicit condition on the mixing parameter $p$ that ensures $ \rho $ remains PPT entangled and detected by the same witness operator $W$. Additionally, we apply our PPT condition to these mixtures, yielding further criteria for verifying positivity under partial transposition.\\
Finally, we consider a novel class of states as sums of tensor products involving Bell states and PPT entangled states of the aforementioned form. This class is particularly relevant to quantum cryptography, as we demonstrate that it exhibits a positive key rate $K_D$, making it viable for secure key distribution protocols \cite{Devetak2005}.\\
The remaining paper is organised as follows: In 
section~\ref{sec-2:Established_Results}, we present several well-known results that will be utilised in the following sections. Section~\ref{sec-3:Condition_On_p2_For_PPT} presents a moment-based PPT criterion to detect PPT states. In section~\ref{sec-4:Lower_Bound_of_p2}, we give the lower bound of the second order moment \( p_2\) of the partial transpose of a PPT state $\rho$. Section~\ref{sec-5:Identification_of_PPTES} explores bound entanglement in the convex mixtures of separable and PPT entangled states. In section~\ref{sec-6:Application}, we propose a novel class of states constructed as normalised sums of tensor products of Bell states and PPT entangled states. These states demonstrate a positive key rate, rendering them valuable for applications in quantum cryptography. Lastly, we conclude in section~\ref{sec-7:Conclusion} with discussions and open questions.
\section{A few established results}
\label{sec-2:Established_Results} 
In this section, we state a few well-established results that will be used in the subsequent section.\\
\textbf{Result-1 \cite{Wolkowicz1980}:} If $A$ be a complex matrix of order $n$   with real eigenvalues $\lambda(A)$, then the lower and upper bounds of the minimum eigenvalue of $A$ are given by
\begin{eqnarray}
m - s\sqrt{n - 1} \leq \lambda_{min}(A) \leq m - \frac{s}{\sqrt{n - 1}}
\label{result1}
\end{eqnarray}
where $m = \frac{\operatorname{Tr}[A]}{n}$, and $s^2 = \frac{\operatorname{Tr}[A^2]}{n} - m^2$.\\
\textbf{Result-2 \cite{elben2020} :} If a bipartite system described by the density operator $\rho_{AB}$, which belongs to the set of positive partial transposed (PPT) states and $p_{2}$ and $p_{3}$ denote the second and third moment of the partially transposed state $\rho_{AB}^{T_{B}}$ then for all PPT states, the following inequality holds:
\begin{eqnarray}
p_{2}^{2}\leq p_{3}
\label{result2}
\end{eqnarray}
\textbf{Result-3 \cite{coope} :} If $A$ and $B$ are positive semidefinite operator then
\begin{eqnarray}
(Tr[AB])^{\frac{1}{2}} \leq \frac{1}{2}(Tr[A]+Tr[B])
\label{result3}
\end{eqnarray}
This result was conjectured by Bellman \cite{bellman} and proved by Neudecker \cite{neudecker} and Yang \cite{yang} independently.\\
\textbf{Result-4 \cite{coope} :} For any two positive semidefinite matrices $A$ and $B$ of the same order, we have
\begin{eqnarray}
Tr(AB)\leq Tr(A)Tr(B)
\label{result4}
\end{eqnarray}
\textbf{Result-5 \cite{anu} :} For any two $n \times n$ Hermitian matrices $A$ and $B$, the following result holds
\begin{eqnarray}
\lambda_{min}(A)Tr(B)\leq Tr(AB)\leq \lambda_{max}(A)Tr(B)
\label{result5}
\end{eqnarray}
\section{Detecting PPT States using the second-order moment of the partial transposition operation}
\label{sec-3:Condition_On_p2_For_PPT}
The role of low-order moments of the quantum state in characterising separability has been noted in several foundational works. In \cite{Zyczkowski1998} and 
\cite{Gurvits2002}, it has been shown that the second order moment of the state provides tight upper bound on the state space, for the largest ball of separable and absolutely separable states. A similar result is given in \cite{adhikari2021} for two-qubit zero discord state $\rho_{ZD}$, which satisfy,
\[
Tr(\rho_{ZD}^2)\leq \min\left( \frac{1}{2}+2|\vec{r_1}|^2, \frac{1}{2}+2|\vec{r_2}|^2\right)
\]
where $\vec{r_i}, i=1,2$ are bloch vectors.
Elben et. al. \cite{elben2020} and X-D Yu et. al. \cite{Yu2021} provides moment based entanglement detection criteria based on the second and third order moments of the partially transposed states. So, these results illustrate that even simple low-order moments of the state or its partial transposed state can provide non-trivial information about separability or entanglement.\\
This section aims to provide a moment-based criterion to detect PPT states. PPT criterion introduced by Peres and Horodecki can also detect PPT states, but the problem with the partial transposition operation is that it cannot be implemented in the laboratory. So, we have adopted a moment-based criterion, which may be applicable in the real setup to detect PPT states. Our finding is that there exists a value (dependent only on the dimension of the system) of the second-order moment of partial transposition of the given density matrix, below which the density matrix under probe is a PPT state. This condition is necessary, but not sufficient. We may note here that the given criterion can detect PPT states, but it is unable to discriminate between the separable states and PPT entangled states.
Let us now state the necessary condition for a quantum state to be a PPT state.
\begin{theorem}
\label{thm: moment_criteria_for_PPT}
Let us consider a $d_{1} \otimes d_{2}$ dimensional system expressed by the density operator $\rho_{AB}$, where the subsystems $A$ and $B$ described by the Hilbert spaces $H_{A}$ and $H_{B}$ respectively and $\rho_{AB}^{T_B}$ is the partial transposition of the density matrix $\rho_{AB}$. Suppose that  
$p_{2}(\rho_{AB}^{T_B})$ denote the second order moment of the partial transposition of $\rho_{AB}$ i.e. $p_{2}(\rho_{AB}^{T_B})=Tr\left[(\rho_{AB}^{T_B})^2\right]$. The necessary condition that if $p_{2}(\rho_{AB}^{T_B}) \leq \frac{1}{d_{1}d_{2}-1}$, then $\rho_{AB}$ is a PPT state.
\end{theorem}
Proof: To prove the necessary condition, we will use Result-1. In (\ref{result1}), we replace the complex matrix $A$ of order $n$ with the partial transposition of the density matrix $\rho_{AB}^{T_B}$ of order $d_{1}d_{2}$. Therefore, (\ref{result1}) reduces to
\begin{eqnarray}
\begin{split}
\frac{1}{d_{1}d_{2}}-\frac{\sqrt{(p_{2}(\rho_{AB}^{T_B})d_{1}d_{2}-1)(d_{1}d_{2}-1)}}{d_{1}d_{2}}\leq \lambda_{min}(\rho_{AB}^{T_B})  \\ \leq
\frac{1}{d_{1}d_{2}}-\frac{1}{d_{1}d_{2}}\sqrt{\frac{p_{2}(\rho_{AB}^{T_B})d_{1}d_{2}-1}{d_{1}d_{2}-1}}
\label{ineq1}
\end{split}
\end{eqnarray}
From (\ref{ineq1}), we can say that $\lambda_{min}(\rho_{AB}^{T_B})\geq 0$ if $p_{2}(\rho_{AB}^{T_B})$ satisfies the inequality
\begin{eqnarray}
\frac{1}{d_{1}d{2}}-\frac{\sqrt{(p_{2}(\rho_{AB}^{T_B})d_{1}d_{2}-1)(d_{1}d_{2}-1)}}{d_{1}d_{2}}\geq 0
\label{ineq2}
\end{eqnarray}
Simplifying (\ref{ineq2}), we get
\begin{eqnarray}
p_{2}(\rho_{AB}^{T_B})\leq \frac{1}{d_{1}d_{2}-1}
\label{neccond}
\end{eqnarray}
Therefore, if the inequality (\ref{neccond}) holds then the $d_{1}\otimes d_{2}$ dimensional state $\rho_{AB}$ is a PPT state. The above result (\ref{neccond}) is stronger than Result-2, as we can use here the inequality (\ref{neccond}) to detect PPT states. Moreover, we can analyze that if the inequality (\ref{neccond}) holds for any two-qubit system described by the density operator $\rho_{AB}$ then the state $\rho_{AB}$ must be separable, but this conclusion doesn't hold for higher-dimensional systems, as there exist PPT entangled states also. Let us verify the result given in (\ref{neccond}) with a few examples taken from the 4-dimensional and 9-dimensional system.\\
\begin{example}
\label{Exm:Thm_1_example_1}
Consider the following quantum state described by the $2\otimes 2$-dimensional density operator $\rho_{AB}^{(1)}$
\[
\rho_{AB}^{(1)}=
\frac{1}{100}
\begin{bmatrix}
27 & 0 & 8 & 4 \\
0 & 13 & -13 & 1 \\
8 & -13 & 32 & -4 \\
4 & 1 & -4 & 28
\end{bmatrix}
\]
The second order moment of $(\rho_{AB}^{(1)})^{T_{B}}$ is denoted by $p_{2}((\rho_{AB}^{(1)})^{T_{B}})$ and is given by
\begin{eqnarray}
p_{2}((\rho_{AB}^{(1)})^{T_{B}})=0.3238 \leq \frac{1}{(2)^2-1}=\frac{1}{3}
\end{eqnarray}
Therefore, the inequality (\ref{neccond}) is verified and thus from theorem (\ref{thm: moment_criteria_for_PPT}), we can say that $\rho_{AB}^{(1)}$ represent a $PPT$ state. In this case, we can certainly say that the state $\rho_{AB}^{(1)}$ is a separable state as it belongs to $2\otimes 2$ system.\\
\end{example}
\begin{example}
\label{Exm:Thm_1_example_2} In $2\otimes 3$ system, let us  consider the following state
\[
\rho_{AB}^{(2)}=
\frac{1}{100}
\begin{bmatrix}
9 & -4 & -3 & -1 & -3 & 3 \\
-4 & 21 & 0 & 2 & -1 & -1 \\
-3 & 0 & 20 & 0 & 6 & -2 \\
-1 & 2 & 0 & 13 & -1 & 0 \\
-3 & -1 & 6 & -1 & 17 & 4 \\
3 & -1 & -2 & 0 & 4 & 20
\end{bmatrix}
\]
For the quantum state described by the density operator $\rho_{AB}^{(2)}$, the value of $p_{2}((\rho_{AB}^{(2)})^{T_{B}})$ is given by
\begin{eqnarray}
p_{2}((\rho_{AB}^{(2)})^{T_{B}})=0.1994<\frac{1}{2\times 3-1}=0.2
\end{eqnarray}
Therefore, the inequality (\ref{neccond}) is also verified by the quantum state $\rho_{AB}^{(2)}$ and by theorem (\ref{thm: moment_criteria_for_PPT}), we conclude that $\rho_{AB}^{(2)}$ is a $PPT$ state. It also represents a separable state, as Peres-Horodecki criterion states that a  $2\otimes2$ dimensional and $2\otimes3$ dimensional states are PPT if and only if they are separable states.\\
\end{example}
\begin{example}
\label{Exm:Thm_1_example_3} Let us now consider the state $\rho_{AB}^{(3)}$ in $3\otimes 3$ dimensional system as
\[
\rho_{AB}^{(3)}=
\begin{bmatrix}
\frac{1}{8} & 0 & 0 & 0 & \frac{1}{48} & 0 & 0 & 0 & \frac{1}{48} \\
0 & \frac{5}{48} & 0 & 0 & 0 & 0 & 0 & 0 & 0 \\
0 & 0 & \frac{5}{48} & 0 & 0 & 0 & 0 & 0 & 0 \\
0 & 0 & 0 & \frac{5}{48} & 0 & 0 & 0 & 0 & 0 \\
\frac{1}{48} & 0 & 0 & 0 & \frac{1}{8} & 0 & 0 & 0 & \frac{1}{48} \\
0 & 0 & 0 & 0 & 0 & \frac{5}{48} & 0 & 0 & 0 \\
0 & 0 & 0 & 0 & 0 & 0 & \frac{5}{48} & 0 & 0 \\
0 & 0 & 0 & 0 & 0 & 0 & 0 & \frac{5}{48} & 0 \\
\frac{1}{48} & 0 & 0 & 0 & \frac{1}{48} & 0 & 0 & 0 & \frac{1}{8}
\end{bmatrix}
\]
Following the earlier examples, it can be easily verified that the state $\rho_{AB}^{(3)}$ satisfies the inequality (\ref{neccond}) as $p_{2}((\rho_{AB}^{(3)})^{T_{B}})=0.114583$ which is less than $0.125$. Therefore, by theorem (\ref{thm: moment_criteria_for_PPT}), we can infer that the state $\rho_{AB}^{(3)}$ represents only a $PPT$ state, but in this case, we cannot discriminate between the separable state and the PPT entangled state.\\
\end{example}
\begin{example}
\label{Exm:Thm_1_example_4} Let us now consider the state $\rho_{AB}^{(4)}$ in $3\otimes 3$ dimensional system,
\[
\rho_{AB}^{(4)}=
\begin{bmatrix}
\frac{3}{25} & 0 & 0 & 0 & 0 & 0 & 0 & 0 & 0 \\
0 & \frac{2}{25} & 0 & 0 & 0 & 0 & 0 & 0 & 0 \\
0 & 0 & \frac{13}{100} & 0 & 0 & 0 & 0 & 0 & 0 \\
0 & 0 & 0 & \frac{14}{100} & 0 & 0 & 0 & 0 & 0 \\
0 & 0 & 0 & 0 & \frac{3}{25} & 0 & 0 & 0 & \frac{1}{25} \\
0 & 0 & 0 & 0 & 0 & \frac{2}{25} & 0 & -\frac{1}{20} & 0 \\
0 & 0 & 0 & 0 & 0 & 0 & \frac{2}{25} & 0 & 0 \\
0 & 0 & 0 & 0 & 0 & -\frac{1}{20} & 0 & \frac{13}{100} & 0 \\
0 & 0 & 0 & 0 & \frac{1}{25} & 0 & 0 & 0 & \frac{3}{25}
\end{bmatrix}
\]
In this example also it can be easily verified that $\rho_{AB}^{(4)}$ satisfies the inequality (\ref{neccond}) as $p_{2}((\rho_{AB}^{4})^{T_{B}})=0.124$ which is less than $0.125$. Therefore, by theorem (\ref{thm: moment_criteria_for_PPT}), we can infer that the state $\rho_{AB}^{(4)}$ represents a $PPT$ state, and this can also be verified by the Peres-Horodecki $PPT$ criterion. Later, we shall show that the state $\rho_{AB}^{(4)}$ is a PPT entangled state.\\
\end{example}
\begin{corollary}
\label{cor:neccond}
If a $d_{1}\otimes d_{2}$ dimensional bipartite quantum state is a negative partial transpose entangled state described by the density operator  $\rho_{AB}^{NPTES}$, then the following inequality holds
\begin{eqnarray}
p_{2}((\rho_{AB}^{NPTES})^{T_B}) > \frac{1}{d_{1}d_{2}-1}
\label{neccondcor}
\end{eqnarray}
\end{corollary}
\begin{example}
\label{Exm:Cor_2_example_1} Let us consider the following $2\otimes 3 $ dimensional quantum state,
\[
\rho_{AB}^{NPTES}=
\begin{bmatrix}
0.19 & 0 & 0 & 0 & 0 & 0.13 \\
0 & 0.15 & 0.11 & 0 & 0 & 0 \\
0 & 0.11 & 0.18 & 0.02 & 0 & 0 \\
0 & 0 & 0.02 & 0.16 & 0.09 & 0 \\
0 & 0 & 0 & 0.09 & 0.13 & 0 \\
0.13 & 0 & 0 & 0 & 0 & 0.19
\end{bmatrix}
\]
This is an NPTES quantum state as the minimum eigenvalue of the partial transpose of $\rho_{AB}^{NPTES}$ is $-0.022$, the value of $p_2$ is $0.2446$, which is greater than $\frac{1}{d_1 d_2-1}=0.2$. Hence, corollary (\ref{cor:neccond}) is verified.
Let us now consider an example in $3 \otimes 3$ dimensional system.
\end{example}
 \begin{example}
\label{Exm:Cor_2_example_2} Consider the following NPTES quantum state,
\[
\begin{bmatrix}
0.09 & 0.05 & 0.02 & 0 & 0.01 & 0 & 0.02 & 0.03 & 0.04 \\
0.05 & 0.13 & 0.03 & 0.02 & 0.06 & 0.04 & 0.01 & 0 & 0.02 \\
0.02 & 0.03 & 0.10 & 0 & 0 & 0.05 & 0.05 & 0 & 0.03 \\
0 & 0.02 & 0 & 0.10 & 0.05 & 0.04 & 0.02 & 0.04 & 0 \\
0.01 & 0.06 & 0 & 0.05 & 0.14 & 0.04 & 0 & 0.05 & 0.04 \\
0 & 0.04 & 0.05 & 0.04 & 0.04 & 0.10 & 0 & 0 & 0 \\
0.02 & 0.01 & 0.05 & 0.02 & 0 & 0 & 0.10 & 0.05 & 0.01 \\
0.03 & 0 & 0 & 0.04 & 0.05 & 0 & 0.05 & 0.13 & 0.06 \\
0.04 & 0.02 & 0.03 & 0 & 0.04 & 0 & 0.01 & 0.06 & 0.11 \\
\end{bmatrix}
\]
For this NPTES state, $p_2=0.1872>\frac{1}{d_1 d_2-1}=\frac{1}{8}$. Hence, corollary (\ref{cor:neccond}) is verified.
\end{example}
\section{Condition on the bloch vector for $d \otimes d$ dimensional PPT states}
In this section, we express the PPT condition given in theorem (\ref{thm: moment_criteria_for_PPT}) in terms of the bloch vector of the underlying quantum state. To achieve this, we use the bloch vector representation of the bipartite two-qudit quantum state described by the density operator $\rho_{AB}$. Therefore,  we can expand the density matrix   $\rho_{AB}$ with respect to the generalised Gell–Mann matrix basis as \cite{kimura2003,byrd2003, bertlmann2008, Gamel2016} .
\begin{equation}
\label{eqn:rho_in_bloch_vector_form}
    \rho_{AB}=\frac{1}{d^2}I_{d^2}+\frac{1}{2}\sum_{k=1}^{d^2-1} r_k\Lambda_k
\end{equation}
where $I_{d^2}$ is the identity matrix of dimension $d^2$, $\vec{r}=(r_1, ..., r_{d^2-1}) \in \mathbb{R}^{d^2-1}$ be the bloch vector such that $|\vec{r}|\leq \sqrt{\frac{2(d^2-1)}{d^2}}$ and $\vec{\Lambda}=(\Lambda_1,...,\Lambda_{d^2-1})$ be the vector of generalised Gell-Mann matrices (traceless generators of $SU(d^2))$. Thus, we have
$Tr(\Lambda_k)=0$ for $k= 1,..., d^2-1$ and
$Tr(\Lambda_k \Lambda_m)=2\delta_{km}$.\\ Using the PPT criterion given in theorem (\ref{thm: moment_criteria_for_PPT}) and the expansion (\ref{eqn:rho_in_bloch_vector_form}) for $\rho_{AB}^{T_{B}}$, where $T_{B}$ denote the partial transposition wih respect to he subsystem $B$, we obtain a condition on the bloch vector that describe the structure of the PPT states. Therefore, we are successful in sketching the geometry of the PPT states through the condition on the   bloch vector obtained after applying the partial transposition operation on the given state. Now, we are in a position to state this condition, which is given below,
\begin{theorem}
    \label{thm:bloch_sphere_PPT_condition}
A bipartite quantum state described by the density operator $\rho_{AB}$, in $d \otimes d$ dimensional system, represents a PPT state if the Bloch vector $\vec{r}$ in the expansion of $\rho_{AB}$ in the basis of generalised Gell-Mann matrices satisfies 
\[|\vec{r}|\leq \sqrt{\frac{2}{d^2(d^2-1)}}\]
\end{theorem}
\textbf{Proof:} Let us consider a $d \otimes d$ dimensional bipartite state described by the density operator $\rho_{AB}$. By applying the partial transposition operation on both sides of (\ref{eqn:rho_in_bloch_vector_form}), $\rho_{AB}^{T_{B}}$ can be represented as
\begin{equation}
\rho_{AB}^{T_B}=\frac{1}{d^2}I_{d^2}+\frac{1}{2}\sum_{k=1}^{d^2-1} r_k\Lambda^{T_B}_k
\end{equation}
where $\Lambda^{T_B}_k, k=1,1,2,...,d^2-1$ denote the partially transposed generalised Gell-Mann matrices with respect to the second subsystem.\\
Now, we have
\begin{align}
    (\rho_{AB}^{T_B})^2&=\frac{1}{d^4}I_{d^2}+\frac{1}{d^2}\sum_{k=1}^{d^2-1}r_k \Lambda^{T_B}_k+\frac{1}{4}\sum_{k,m=1}^{d^2-1}r_k r_m \Lambda^{T_B}_k \Lambda^{T_B}_m
\end{align}
Therefore, 
\begin{align}
p_2(\rho_{AB}^{T_B})=\mathrm{Tr}[(\rho^{T_B})^2]
 &=\frac{1}{d^2}+\frac{1}{2} \sum_{k=1}^{d^2-1} r_{k}^{2}=\frac{1}{d^2} +\frac{1}{2}|\vec{r}|^2 
\label{blochcond}
\end{align}
Now, if we impose the PPT condition given in theorem (\ref{thm: moment_criteria_for_PPT}) for $d\otimes d$ dimensional system, then we have 
\begin{align}
p_2(\rho_{AB}^{T_B})\leq \frac{1}{d^2-1}
\label{pptcond1}
\end{align}
Combining the conditions (\ref{blochcond}) and (\ref{pptcond1}), we get
\begin{align}
|\vec{r}|\leq \sqrt{\frac{2}{d^2(d^2-1)}}
\label{blochcond2}
\end{align}
Therefore, we can conclude that if the bloch vector $\vec{r}$ corresponding to $\rho_{AB}$ satisfies $|\vec{r}|\leq \sqrt{\frac{2}{d^2(d^2-1)}}$ then the state $\rho_{AB}$ represent a PPT state.\\
\textbf{Example 1:} Let us consider the two-qubit Werner state 
\begin{equation}
    \rho_W=\frac{1-x}{4}I+xS,\qquad 0\leq x\leq 1
\end{equation}
Where S is given by
\[
S = \frac{1}{2}
\begin{bmatrix}
0 & 0 & 0 & 0 \\
0 & 1 & -1 & 0 \\
0 & -1 & 1 & 0 \\
0 & 0 & 0 & 0
\end{bmatrix}.
\]
We can re-express $\rho_W$ in the generalised Gell-Mann matrix basis as \cite{kimura2003},
\begin{align}
\rho_W=\frac{1}{4}I_4+\frac{1}{2}\vec{r}. \vec{\Lambda}
\end{align}
where $\Lambda_i$'s are generalised Gell-Mann matrices for four dimensional system \cite{bertlmann2008} and $\vec{r}$ is given by
\[
\vec{r}=(0,0,0, -x, 0,0,0,0,0,0,0,0, -\frac{x}{2}, -\frac{x}{2\sqrt{3}},\frac{x}{\sqrt{6}})
\]
The operator $\rho_W$ is positive semi-definite if 
\begin{align}
\label{eqn:positive_cond}
    |\vec{r}|&\leq \sqrt{\frac{3}{2}} \Rightarrow x\leq 1
\end{align}
Moreover, the operator $\rho_W^{T_B}$ is positive semi-definite if 
\begin{align}
\label{eqn:PPT_cond}
    |\vec{r}|&\leq \sqrt{\frac{1}{6}} \Rightarrow x\leq \frac{1}{3}
\end{align}
Therefore, \ref{eqn:PPT_cond} tells us that $\rho_W$ is a PPT i.e. a separable state when $x\leq \frac{1}{3}$, which matches with the standard result for Werner state.\\
\section{Moments Based Entanglement detection criterion using Newton's identities}
Let $\rho_{AB}$ be a quantum state acting on $d_1 d_2$ dimensional system, where $d_1$ and $d_2$  are the dimensions of the subsystems $A$ and $B$, respectively.
Let $det(\rho_{AB}^{T_B}-\lambda I_{d_1 d_2})$ be the characteristic polynomial of $\rho_{AB}^{T_B}$. The characteristic equation of $\rho_{AB}^{T_B}$ is given by 
\begin{equation}
    \lambda^{d_1 d_2}-S_1 \lambda^{d_1 d_2-1}+S_2 \lambda^{d_1 d_2-2}-...+(-1)^{d_1 d_2}S_{d_1 d_2}=0
\end{equation}
where $S_k$ are the symmetric functions given as \cite{horn1990} 
\begin{equation}
    S_k=\sum_{1\leq i_1\leq... \leq i_k\leq d_1 d_2} \prod_{j=1}^k \lambda_{i_j}
\end{equation}
where $\lambda_{i}$'s are the eigenvalues of $\rho_{AB}^{T_B}$.
We can write $S_k$'s in terms of the functions of $p_k\left(\rho_{AB}^{T_B}\right), k=1,2,...,d_1 d_2$ as \cite{byrd2003,Homa2023}\\
\[
S_1=p_1(\rho_{AB}^{T_B})=1
\]
\begin{align*}
 S_2 &=\frac{1}{2}\left[1-p_2\left(\rho_{AB}^{T_B}\right)\right]
\end{align*}
and 
\begin{equation}
\label{eqn:newton_iden}
\begin{split}
 S_k&=\frac{1}{k}\Bigg[S_{k-1}-p_2\left(\rho_{AB}^{T_B}\right) S_{k-2}+...\\
 &\quad+(-1)^{d_1 d_2-1}p_{d_1 d_2}\left(\rho_{AB}^{T_B}\right) S_{k-d_1 d_2}+...\\
 &\quad+(-1)^{k-2}p_{k-1}\left(\rho_{AB}^{T_B}\right)+(-1)^{k-1}p_k\left(\rho_{AB}^{T_B}\right)\Bigg] 
 \end{split}
\end{equation}
where $p_k=Tr(\rho_{AB}^{T_B})^k$. The identities given by (\ref{eqn:newton_iden}) for $k=1,2,3,...,d_1 d_2$ are call Newton's identities or Newton–Girard identities. By Descartes' rule of signs, we can say that  iff $S_k\geq 0$ for all $k$ then $\rho_{AB}^{T_B}$ is positive semi-definite. Hence, $\rho_{AB}$ is a PPT state.\\
If at least one of the $S_k$'s are less than zero then we can say that $\rho_{AB}^{T_B}$ has at least one negative eigenvalue and hence $\rho_{AB}$ must be entangled.
In particular, let us consider $S_3$ is negative i.e. $S_3<0$, and it implies the following inequality,
\begin{align}
\begin{split}
\label{eqn:ent_cond_Newton}
    &\frac{1}{3}\left[S_2-p_2(\rho_{AB}^{T_B})+p_3(\rho_{AB}^{T_B})\right] <0\\
   \Rightarrow
    &p_2(\rho_{AB}^{T_B})>\frac{2}{3}p_3(\rho_{AB}^{T_B})+\frac{1}{3}
    \end{split}
\end{align}
Therefore, if $ p_2(\rho_{AB}^{T_B})>\frac{2}{3}p_3(\rho_{AB}^{T_B})+\frac{1}{3}$ holds then the state $\rho_{AB}$ represents a negative partial transpose entangled state.\\
Let us consider the following example.\\
\textbf{Example:} Consider the following $2\otimes 3$ dimensional state.
\[
\rho_{AB}=
\left[
\begin{array}{cccccc}
 0.11 & 0.12 & 0.05 & 0.01 & -0.04 & 0.1 \\
 0.12 & 0.36 & 0.09 & 0.05 & -0.06 & 0.25 \\
 0.05 & 0.09 & 0.06 & -0.03 & 0.03 & 0.07 \\
 0.01 & 0.05 & -0.03 & 0.08 & -0.06 & 0.03 \\
 -0.04 & -0.06 & 0.03 & -0.06 & 0.17 & 0. \\
 0.1 & 0.25 & 0.07 & 0.03 & 0. & 0.22 \\
\end{array}
\right]
\]
The value of $p_2(\rho_{AB}^{T_B})$ is 0.462 and the value of $p_3(\rho_{AB}^{T_B})$ is 0.148846. Therefore, the condition given in (\ref{eqn:ent_cond_Newton}) is satisfied as $p_2(\rho_{AB}^{T_B})-\frac{2}{3}p_3(\rho_{AB}^{T_B})-\frac{1}{3}=0.0294$. Therefore, $\rho_{AB}$ must be an entangled state. This result is correct as the minimum eigenvalue of $\rho_{AB}^{T_B}$ is $-0.2168$, therefore the $\rho_{AB}$ is entangled state. 
\section{Lower bound of the second order moment of partial transposition operation}
\label{sec-4:Lower_Bound_of_p2}
In spite of having the upper bound of the second-order moment of the partial transposition operation, we still lack the lower bound of the same. Therefore, it is natural to ask about the lower bound of $p_{2}$. The trivial answer to the above-asked question is zero, but in this section, we are in search of a non-trivial lower bound of $p_{2}$. Our investigation suggests that the non-trivial lower bound of $p_{2}$ can be calculated for the set of PPT states, that is, if it is known that the state under investigation is PPT, then we can derive the non-trivial lower bound of the second-order moment of the partial transposition of the given state.
\begin{theorem}
\label{thm2}
Let us consider a $d_{1} \otimes d_{2}$ dimensional PPT state $\rho_{AB}$, where the subsystems $A$ and $B$ described by the Hilbert spaces $H_{A}$ and $H_{B}$ respectively. If $p_{2}$ and $p_{3}$ denote the second and third ordered moment of $\rho_{AB}^{T_{B}}$ i.e. if $p_{2}=Tr\left[(\rho_{AB}^{T_B})^2\right]$ and $p_{3}=Tr\left[(\rho_{AB}^{T_B})^3\right]$ then $p_{2}$ and $p_{3}$ satisfies the inequality
\begin{eqnarray}
2\sqrt{p_{3}}-1 \leq p_{2} \leq \sqrt{p_{3}}
\label{ineq3}
\end{eqnarray}
\end{theorem}
\textbf{Proof:} To derive the non-trivial lower bound of $p_{2}$, we use Result-3 in which the positive semidefinite operators $A$ and $B$ are replaced by $\rho_{AB}^{T_{B}}$ and $(\rho_{AB}^{T_{B}})^{2}$. Here, in this case the above replacements are possible, since the given states are PPT and thus the matrices $\rho_{AB}^{T_{B}}$ and $(\rho_{AB}^{T_{B}})^{2}$ can be considered as density matrices and thus a positive semidefinite matrices.\\
With a suitable modification in Result-3, we get
\begin{align}
\begin{split}
(Tr[\rho_{AB}^{T_{B}}.((\rho_{AB}^{T_{B}})^{2})])^{\frac{1}{2}}\leq& \frac{1}{2}[Tr[\rho_{AB}^{T_{B}}+Tr[(\rho_{AB}^{T_{B}})^{2}]]
\end{split}
\label{ineq4}
\end{align}
Inequality (\ref{ineq4}) can be expressed in terms of $p_{2}$ and $p_{3}$ as
\begin{eqnarray}
\sqrt{p_{3}} \leq \frac{1}{2}[1+p_{2}]
\label{ineq5}
\end{eqnarray}
Simplifying (\ref{ineq5}), we write the lower bound of $p_{2}$ in terms of $p_{3}$ as
\begin{eqnarray}
2\sqrt{p_{3}}-1 \leq p_{2}
\label{lb}
\end{eqnarray}
Thus, combining the inequalities (\ref{result2}) and (\ref{lb}), we get
\begin{eqnarray}
2\sqrt{p_{3}}-1 \leq p_{2} \leq \sqrt{p_{3}}
\label{ublb}
\end{eqnarray}
Hence, if the given state is a PPT state, then the above inequality (\ref{ublb}) holds. Now, it is worth investigating whether there exists any PPT quantum state for which the lower bound $2\sqrt{p_{3}}-1$ is positive and also satisfies the inequality (\ref{ublb}). We now show that the above statement is indeed correct with a few examples, which are given below.\\
\begin{example}
\label{Exm:Thm_3_example_1} Let us consider the following bipartite PPT quantum state described by the density operator $\rho_{AB}^{(4)}$
\[
\rho_{AB}^{(4)}=
\begin{bmatrix}
0.35 & -0.05 & -0.26 & -0.01 \\
-0.05 & 0.26 & -0.10 & 0 \\
-0.26 & -0.10 & 0.34 & 0.06 \\
-0.01 & 0 & 0.06 & 0.05
\end{bmatrix}, 
\]
It can be observed that the value of $p_{2}((\rho_{AB}^{(4)})^{T_{B}})$ is $0.4758$, which is greater than $\frac{1}{3}$ and thus our criterion (\ref{neccond}) does not detect that the state $\rho_{AB}^{(4)}$ is a PPT state but it can be proved from other PPT criterion that the state $\rho_{AB}^{(4)}$ is indeed a PPT state. The value of the third-order moment of the partial transposition of the state $\rho_{AB}^{(4)}$ is found to be 0.2694, i.e. $p_3((\rho_{AB}^{(4)})^{T_{B}})=0.2694$. Therefore, the lower bound of $p_{2}((\rho_{AB}^{(4)})^{T_{B}})$ can be calculated from (\ref{ublb}) and is given by $2\sqrt{p_3}-1= 0.038$. Thus, it can be easily seen that the inequality (\ref{ublb}) is verified for the PPT state described by the density operator $\rho_{AB}^{(4)}$.
\end{example}
\begin{example}
\label{Exm:Thm_3_example_2}
\begin{widetext}
Another PPT quantum state in $2\otimes 3$ dimensional system described by the density operator $\rho_{AB}^{(5)}$ as
\[
\rho_{AB}^{(5)}=
\begin{bmatrix}
0.0855788 & -0.0130138 & -0.0634194 & -0.0602343 & 0.0151165 & 0.0556449 \\
-0.0130138 & 0.0319954 & 0.0319794 & 0.00361884 & -0.0293307 & -0.0151244 \\
-0.0634194 & 0.0319794 & 0.326903 & 0.075471 & 0.00431698 & -0.239706 \\
-0.0602343 & 0.00361884 & 0.075471 & 0.0891845 & -0.0445194 & -0.0865549 \\
0.0151165 & -0.0293307 & 0.00431698 & -0.0445194 & 0.100965 & 0.0767125 \\
0.0556449 & -0.0151244 & -0.239706 & -0.0865549 & 0.0767125 & 0.365373
\end{bmatrix}
\]  
\end{widetext}
We find that $p_{2}((\rho_{AB}^{(5)})^{T_{B}})=0.45046$ and $p_{3}((\rho_{AB}^{(5)})^{T_{B}})=0.266987$, so in this case, our criterion (\ref{neccond}) fails to detect $\rho_{AB}^{(5)}$ as a PPT quantum state but one can easily verify the inequality given in (\ref{ublb}).\\  
Now, it may be noted that the entangled state can be detected by the contrapositive statement of $Theorem-2$, which is stated below:\\ 
\end{example}
 \begin{corollary} If $d_{1}\otimes d_{2}$ dimensional bipartite quantum state described by the density operator $\rho_{AB}$ and if either
$p_{2}(\rho_{AB}^{T_{B}}) < 2\sqrt{p_{3}(\rho_{AB}^{T_{B}})}-1
$
or 
$
p_{2}(\rho_{AB}^{T_{B}}) > \sqrt{p_{3}(\rho_{AB}^{T_{B}})
}$ holds then the state $\rho_{AB}$ is a NPT entangled state.
\end{corollary}
\section{Identification of a quantum state as a PPT entangled state}
\label{sec-5:Identification_of_PPTES}
In this section, we defined a family of $d_{1}\otimes d_{2}$ dimensional quantum states and derived a condition by which we can identify it as a PPT state. We use the results discussed in the previous sections to accomplish this task. Once we detect that the newly defined state is PPT under certain conditions, we apply a witness operator to find out whether the PPT state represents a family of PPT entangled states.\\  
To achieve the task, we define a family of $d_{1}\otimes d_{2}$ dimensional quantum states by considering the convex combination of a separable state and a PPT entangled state. Mathematically, the defined state can be expressed as 
\begin{equation}
 \rho_{AB}^{PE}=p\rho_{SEP}+(1-p)\rho_{PPTES},~~ 0\leq p \leq 1
 \label{sepbes}
\end{equation}
where $\rho_{SEP}$ and $\rho_{PPTES}$ denote the bipartite separable and the PPT entangled state respectively, in $d_{1}\otimes d_{2}$ dimensional system.\\
Since $\rho_{AB}^{PE}$ is a convex combination of separable and PPT entangled states so it is very legitimate to investigate whether the density operator $\rho_{AB}^{PE}$ represent a separable state or a PPT entangled state. We now move on to investigate this question and find that we can answer the above-asked question in two steps. In the first step, we will use $Theorem-1$ and verify that $\rho_{AB}^{PE}$ represents a PPT state under certain conditions. But in a higher-dimensional system, the PPT state means that it may represent either a separable state or a PPT entangled state. Therefore, we will proceed towards the second step, where we probe for a witness operator that may detect $\rho_{AB}^{PE}$ as an entangled state. So, if there exists any such witness operator, then combining the above-mentioned two steps, we are able to say that the state $\rho_{AB}^{PE}$ is a PPT entangled state.\\
\textbf{Step-I:} To start with, let us first calculate the second-order moment of the partial transposition of any arbitrary $d_{1} \otimes d_{2}$ dimensional quantum state described by the density operator $\rho_{AB}^{PE}$. Therefore, the second-order moment $p_{2}((\rho_{AB}^{PE})^{T_{B}})$ is given by
\begin{widetext}
\begin{eqnarray}
p_{2}((\rho_{AB}^{PE})^{T_{B}}))=    Tr[((\rho_{AB}^{PE})^{T_{B}}))^2]&=p^2Tr[(\rho_{SEP}^{T_B})^2]+(1-p)^2Tr(\rho_{PPTES}^{T_B})^2+2p(1-p)Tr[\rho_{SEP}^{T_B}.\rho_{PPTES}^{T_B}]\nonumber\\&
\leq p^2 p_{2}(\rho_{SEP}^{T_{B}})+(1-p)^2 p_{2}(\rho_{PPTES}^{T_{B}})+2p(1-p) Tr(\rho_{SEP}^{T_B})Tr(\rho_{PPTES}^{T_B})\nonumber\\&
                   = p^2 p_{2}(\rho_{SEP}^{T_{B}})+(1-p)^2 p_{2}(\rho_{PPTES}^{T_{B}})+2p(1-p)~~~~~~~~~~~~~~~~~~~~~~~~~~~~~~~~~
\label{pe}
\end{eqnarray}
 \end{widetext}
The second and third steps of (\ref{pe}) can be obtained by applying $Result-4$ and using the fact that  $Tr(\rho_{SEP}^{T_B})=1$ and $Tr(\rho_{PPTES}^{T_B})=1$. Therefore, the state $\rho_{AB}^{PE}$ represents a PPT state if the following condition holds
\begin{eqnarray}
p^2 p_{2}(\rho_{SEP}^{T_{B}})+(1-p)^2 p_{2}(\rho_{PPTES}^{T_{B}})+2p(1-p) \leq \dfrac{1}{d_{1}d_{2} - 1}\nonumber\\
\label{cond1}
\end{eqnarray}
\textbf{Step-II:} Once we find the value of the mixing parameter $p$ for which (\ref{cond1}) holds, we proceed towards the next step. In the second step, our task is to identify whether the PPT state $\rho_{AB}^{PE}$ represent a separable state or a PPT entangled state. To probe it, we use the linear witness operator method and assume that there exists a witness operator $W$ that may detect the entangled state $\rho_{PPTES}$. Let
$Tr(W\rho_{SEP})=k_1,\quad k_1 \geq 0$ and $\quad Tr(W\rho_{PPTES})=-k_2,\quad k_2> 0$.\\
Therefore,
 \begin{eqnarray}
Tr(W\rho_{AB}^{PE})&=&p\,Tr[W\rho_{SEP}]+(1-p)\,Tr[W\rho_{PPTES}]\nonumber\\&=&
p \,k_1+(1-p)\, (-k_2)
\label{witness1}
 \end{eqnarray}
If $p \,k_1-(1-p)\, k_2<0$ holds, then $Tr(W\rho_{AB}^{PE})<0$ and hence $W$ detects $\rho_{AB}^{PE}$ as an entangled state and we have the following condition on the parameter $p$, which is given below
 \begin{eqnarray}
  \label{cond2}
0 \leq p<\frac{k_2}{k_1+k_2}
\label{witnesscond}
\end{eqnarray}
 Under the condition (\ref{witnesscond}), the state $\rho_{AB}^{PE}$ is entangled and it is detected by the witness operator $W$.\\
We are now in a position to summarise the above discussion in the form of a theorem, which can be stated as
\begin{theorem}
 \label{thm: condition_for_PPT_for_conv_Comb_rho}
Let us consider a $d_{1}\otimes d_{2}$ dimensional quantum state described by the density operator $\rho_{AB}^{PE} = p\,\rho_{\mathrm{SEP}} + (1 - p)\,\rho_{\mathrm{PPTES}}$ and assume the following two conditions
\begin{eqnarray}
(i)~~ p_2((\rho_{AB}^{PE})^{T_{B}}) \leq \dfrac{1}{d_{1}d_{2} - 1}
\label{cond201}
\end{eqnarray}
\begin{eqnarray}
(ii)~~0 \leq p<\frac{k_2}{k_1+k_2}
\label{cond211}
\end{eqnarray}
where the two real numbers $k_1\geq 0$ and $k_2> 0$ are chosen in such a way that $Tr(W\rho_{SEP})=k_1$ and $Tr(W\rho_{ENT})=-k_2$, $W$ denote the witness operator that can detect $\rho_{PPTES}$. If the parameter $p$ satisfies the conditions (\ref{cond201}) and  (\ref{cond211}) then $\rho_{AB}^{PE}$ represent a PPT entangled state $(PPTES)$.
\end{theorem}
\begin{example}
Let us now consider the convex combination of a separable and a PPT entangled state described by the density operators $\rho_{SEP}$ and $\rho_{PPTES} $ respectively. Therefore, we have the state of the form
\begin{equation}
    \rho_{AB}^{PE}=p\,\rho_{SEP}+(1-p)\,\rho_{PPTES}
\end{equation}
The separable state $\rho_{SEP}$ and the PPT entangled state $\rho_{PPTES}$ may be expressed in the following form:
\[
\rho_{SEP}=
\begin{bmatrix}
\frac{2}{21} & 0 & 0 & 0 & \frac{2}{21} & 0 & 0 & 0 & \frac{2}{21} \\
0 & \frac{a}{21} & 0 & 0 & 0 & 0 & 0 & 0 & 0 \\
0 & 0 & \frac{5-a}{21} & 0 & 0 & 0 & 0 & 0 & 0 \\
0 & 0 & 0 & \frac{5-a}{21} & 0 & 0 & 0 & 0 & 0 \\
\frac{2}{21} & 0 & 0 & 0 & \frac{2}{21} & 0 & 0 & 0 & \frac{2}{21} \\
0 & 0 & 0 & 0 & 0 & \frac{a}{21} & 0 & 0 & 0 \\
0 & 0 & 0 & 0 & 0 & 0 & \frac{a}{21} & 0 & 0 \\
0 & 0 & 0 & 0 & 0 & 0 & 0 & \frac{5-a}{21} & 0 \\
\frac{2}{21} & 0 & 0 & 0 & \frac{2}{21} & 0 & 0 & 0 & \frac{2}{21}\\  
\end{bmatrix},\, a\in [2,3]
\]
\[
\rho_{PPTES}= \frac{1}{3(1+x+\frac{1}{x})}
\begin{bmatrix}
 1 & 0 & 0 & 0 & 1 & 0 & 0 & 0 & 1 \\
 0 & x & 0 & 0 & 0 & 0 & 0 & 0 & 0 \\
 0 & 0 & \frac{1}{x} & 0 & 0 & 0 & 0 & 0 & 0 \\
 0 & 0 & 0 & \frac{1}{x} & 0 & 0 & 0 & 0 & 0 \\
 1 & 0 & 0 & 0 & 1 & 0 & 0 & 0 & 1 \\
 0 & 0 & 0 & 0 & 0 & x & 0 & 0 & 0 \\
 0 & 0 & 0 & 0 & 0 & 0 & x & 0 & 0 \\
 0 & 0 & 0 & 0 & 0 & 0 & 0 & \frac{1}{x} & 0 \\
 1 & 0 & 0 & 0 & 1 & 0 & 0 & 0 & 1 \\
\end{bmatrix}
\]
Where $x$ is a positive real number.\\
Now, a witness operator $W$, which detect $\rho_{PPTES}$~\cite{Bhattacharya2021, stormer1982} can be expressed in the following form
\[
W =\frac{1}{3+3 \alpha ^2}
\begin{bmatrix}
 \alpha ^2 & 0 & 0 & 0 & -\alpha  & 0 & 0 & 0 & -\alpha ^2 \\
 0 & 0 & 0 & 0 & 0 & 0 & 0 & 0 & 0 \\
 0 & 0 & 1 & 0 & 0 & 0 & 0 & 0 & 0 \\
 0 & 0 & 0 & \alpha ^2 & 0 & 0 & 0 & 0 & 0 \\
 -\alpha  & 0 & 0 & 0 & 1 & 0 & 0 & 0 & 0 \\
 0 & 0 & 0 & 0 & 0 & 0 & 0 & -\alpha  & 0 \\
 0 & 0 & 0 & 0 & 0 & 0 & 0 & 0 & 0 \\
 0 & 0 & 0 & 0 & 0 & -\alpha  & 0 & 1 & 0 \\
 -\alpha ^2 & 0 & 0 & 0 & 0 & 0 & 0 & 0 & \alpha ^2 \\
\end{bmatrix}
\]
We find that for $\alpha=1$, the value of $Tr(W\rho_{PPTES})$ is given by $\frac{3-x}{18(1+x+x^2)}$, which is negative for $x>3$. Therefore, for $\alpha=1$ and $x>3$, $W$ detects $\rho_{PPTES}$.\\
Now, since $\rho_{AB}^{PE}$ is a convex combination of two PPT states, therefore $\rho_{AB}^{PE}$ represents a PPT state for $0 \leq p\leq 1$. For $a=2.5, \alpha=1$ and $x>3$, we obtain $Tr(W\rho_{AB}^{PE})=\frac{(3-x)+p(11x^2+25x-31)}{18(x^2+x+1)}$, which is negative for $0\leq p< \frac{x-3}{11x^2+25x-31}$, where $x>3$. Hence, $\rho_{AB}^{PE}$ is PPT and entangled for $0\leq p< \frac{x-3}{11x^2+25x-31}$, $x>3$, and thus it represents a PPT entangled state for $p \in \left[0, \frac{x-3}{11x^2+25x-31}\right]$, where $x>3$.\\
\end{example}
\section{Application}
\label{sec-6:Application}
The achievement of distillable key rates ($K_D$) represents a fundamental challenge in quantum cryptography, where the extraction of secure keys from shared quantum states determines the practical viability of quantum key distribution (QKD) protocols~\cite{Devetak2005}. While entanglement serves as the primary resource for secure quantum communication, the precise relationship between the nature of entangled states and their cryptographic utility remains an active area of research. Recent work has established sufficient conditions under which certain classes of PPT entangled states yield nonzero $K_D$, thereby expanding the scope of quantum resources available for cryptographic applications~\cite{Devetak2005}. Building upon these theoretical foundations, we have identified a specific class of entangled states that exhibit positive key rates, providing concrete examples of cryptographically useful states that can be generated using a class of PPT entangled states. To illuminate it further, consider a quantum state $\rho_{c}$ of the following form ~\cite{Horodecki2005} \\
\begin{equation}
\begin{split}
\rho_{c} &=\left(\frac{1}{2 Tr(\sigma_0+\sigma_1+\sigma_2+\sigma_3)}\right)( \ket{\phi^+}\bra{\phi^+} \otimes \sigma_0 + \\
    \quad &\ket{\phi^-}\bra{\phi^-} \otimes \sigma_1 + \ket{\psi^+}\bra{\psi^+} \otimes \sigma_2 
       + \ket{\psi^-}\bra{\psi^-} \otimes \sigma_3)
\end{split}
\label{eqn:state_form_for_K_D}
\end{equation}
where $\ket{\phi^{\pm}}$ and $\ket{\psi^{\pm}}$ are Bell states in $\mathbb{C}^2 \otimes \mathbb{C}^2$ and $\sigma_0, \sigma_1, \sigma_2,$ and $\sigma_3$ are PPT entangled states in $\mathbb{C}^d \otimes \mathbb{C}^d$.\\
We should note here that Horodecki et.al.~\cite{Horodecki2005} and D. P. Chi et.al. \cite{Chi2007} considered $\sigma_{i}$'s are positive operators, but in our case, we consider them as a valid density operator. This assumption doesn't affect the result obtained in~\cite{Horodecki2005}. Therefore, we can use their result to calculate the lower bound of the distillable key rate.\\ 
The state $\rho_{c}$ can also be expressed in the following matrix form.
\begin{widetext}
\[
\rho =\left(\frac{1}{2 Tr(\sigma_0+\sigma_1+\sigma_2+\sigma_3)}\right)
\begin{bmatrix}
\sigma_{0} + \sigma_{1} & 0 & 0 & \sigma_{0} - \sigma_{1} \\
0 & \sigma_{2} + \sigma_{3} & \sigma_{2} - \sigma_{3} & 0 \\
0 & \sigma_{2} - \sigma_{3} & \sigma_{2} + \sigma_{3} & 0 \\
\sigma_{0} - \sigma_{1} & 0 & 0 & \sigma_{0} + \sigma_{1}
\end{bmatrix}.
\]
\end{widetext}
Now, let us choose $\sigma_i$'s as follows,
\begin{equation}
\begin{split}
\sigma_0 &= p_0\rho_{SEP}+(1-p_0)\rho_{PPTES}\\
\sigma_1 &= p_1\rho_{SEP}+(1-p_1)\rho_{PPTES}\\
\sigma_2 &= p_2\rho_{SEP}+(1-p_2)\rho_{PPTES}\\
\sigma_3 &= p_3\rho_{SEP}+(1-p_3)\rho_{PPTES}
\end{split}
\end{equation}
where $p_i\in (0,1)$ for $i=0,1,2,3$. We can choose $p_{i}$'s in such a way that $\sigma_0, \sigma_1, \sigma_2,$ and $\sigma_3$ will be PPT entangled states. Such $p_i$'s can be chosen with the help of theorem (\ref{thm: condition_for_PPT_for_conv_Comb_rho}). 
D. P. Chi et.al. \cite{Chi2007} found an expression of distillable key rate in terms of the trace norm of $\sigma_{0}\pm \sigma_{1}$ and $\sigma_{2}\pm \sigma_{3}$, and it is given by
\begin{eqnarray}
K_{D}=1-Q
\label{keyrate}
\end{eqnarray}
where   
\begin{equation}
    Q = -x \log_2 x - y \log_2 y - z \log_2 z - w \log_2 w
\end{equation}
The variables $x, y, z$ and $w$ are given as follows:
\begin{equation}
    \begin{split}    
x = \frac{1}{2} \left( \lVert \sigma_{0} + \sigma_{1} \rVert + \lVert \sigma_{0} - \sigma_{1} \rVert \right)\\
y = \frac{1}{2} \left( \lVert \sigma_{0} + \sigma_{1} \rVert - \lVert \sigma_{0} - \sigma_{1} \rVert \right)\\
z = \frac{1}{2} \left( \lVert \sigma_{2} + \sigma_{3} \rVert + \lVert \sigma_{2} - \sigma_{3} \rVert \right)\\
w = \frac{1}{2} \left( \lVert \sigma_{2} + \sigma_{3} \rVert - \lVert \sigma_{2} - \sigma_{3} \rVert \right)
    \end{split}
\end{equation}
We call $x,y,z,w$ as variables since the values of $x,y,z,w$ will vary for different PPT entangled states $\sigma_{i}$'s. The distillable key rate $K_{D}$ is positive i.e. $K_{D}> 0$ if $Q>0$,
\begin{example}
    \label{exp1:application}
Let us recall the state $\rho_{c}$ in which the states $\sigma_{0}$, $\sigma_{1}$, $\sigma_{2}$, $\sigma_{3}$ can be constructed using $3\otimes 3$ dimensional separable and PPT entangled states described by the density operator $\rho_{SEP}^{(1)}$ and $\rho_{PPTES}^{(1)}$. The states $\rho_{SEP}^{(1)}$ and $\rho_{PPTES}^{(1)}$ are given by
\[
\begin{aligned}
\rho_{\text{SEP}}^{(1)} &= 
\frac{2}{21}\,|00\rangle\langle00|
+ \frac{2.3}{21}\,|01\rangle\langle01|
+ \frac{2.7}{21}\,|02\rangle\langle02| \\[4pt]
&\quad+ \frac{2.7}{21}\,|10\rangle\langle10|
+ \frac{2}{21}\,|11\rangle\langle11|
+ \frac{2.3}{21}\,|12\rangle\langle12| \\[4pt]
&\quad+ \frac{2.3}{21}\,|20\rangle\langle20|
+ \frac{2.7}{21}\,|21\rangle\langle21|
+ \frac{2}{21}\,|22\rangle\langle22| \\[4pt]
&\quad+ \frac{2}{21}\,|00\rangle\langle11|
+ \frac{2}{21}\,|00\rangle\langle22|
+ \frac{2}{21}\,|11\rangle\langle00| \\[4pt]
&\quad+ \frac{2}{21}\,|11\rangle\langle22|
+ \frac{2}{21}\,|22\rangle\langle00|
+ \frac{2}{21}\,|22\rangle\langle11|
\end{aligned}
\]\\
\[
\begin{aligned}
\rho_{\text{PPTES}}^{(1)} &= 
a\,|00\rangle\langle00|
+ c\,|01\rangle\langle01|
+ a\,|02\rangle\langle02| \\[4pt]
&\quad+ a\,|10\rangle\langle10|
+ a\,|11\rangle\langle11|
+ c\,|12\rangle\langle12| \\[4pt]
&\quad+ c\,|20\rangle\langle20|
+ a\,|21\rangle\langle21|
+ a\,|22\rangle\langle22| \\[4pt]
&\quad+ b\,|00\rangle\langle11|
+ b\,|00\rangle\langle22|
+ b\,|11\rangle\langle00| \\[4pt]
&\quad+ b\,|12\rangle\langle21|
+ b\,|21\rangle\langle12|
+ b\,|22\rangle\langle00|
\end{aligned}
\]
where, $a, b, c$ are given by,
\[
a = \frac{1 + \sqrt{5}}{3 + 9\sqrt{5}}, \quad
b = \frac{-2}{3 + 9\sqrt{5}}, \quad
c = \frac{-1 + \sqrt{5}}{3 + 9\sqrt{5}},
\]
We are now in a position to construct the states $\sigma_{0}$, $\sigma_{1}$, $\sigma_{2}$, $\sigma_{3}$ in the following way:
\begin{equation}
\begin{split}
\sigma_{0} &= 0.43 \, \rho_{\text{SEP}}^{(1)} + (1-0.43) \rho_{\text{PPTES}}^{(1)} \\
\sigma_{1} &= 0.45 \, \rho_{\text{SEP}}^{(1)} + (1-0.45) \rho_{\text{PPTES}}^{(1)}\\
\sigma_{2} &= 0.48 \, \rho_{\text{SEP}}^{(1)} + (1-0.48) \rho_{\text{PPTES}}^{(1)}\\
\sigma_{3} &= 0.50 \, \rho_{\text{SEP}}^{(1)} + (1-0.50) \rho_{\text{PPTES}}^{(1)}
\end{split}
\label{sigma1}
\end{equation}
Using theorem (\ref{thm: condition_for_PPT_for_conv_Comb_rho}), we can say that $\sigma_0, \sigma_1, \sigma_2$ and $\sigma_3$ represent four $3\otimes 3$ dimensional PPT entangled states. Using (\ref{sigma1}), we can calculate the value of the variables $x,y,z,w$ and thus the value of $1-Q$
comes out to be 1.00028, which is greater than $0$. Therefore, $K_D(\rho_{c}) > 0$. Thus, the state $\rho_{c}$ given in (\ref{eqn:state_form_for_K_D}) with $\sigma_{i}$'s given in (\ref{sigma1}) is useful in quantum cryptography.
\end{example}
\begin{example}
    \label{exp2:application}
In this example, we consider another quantum state of the form (\ref{eqn:state_form_for_K_D}) with the following $\sigma_{i}$'s $(i=0,1,2,3)$
\begin{equation}
\begin{split}
\sigma_{0} &= 0.45 \, \rho_{\text{SEP}}^{(2)} + (1-0.45) \, \rho_{\text{PPTES}}^{(2)} \\
\sigma_{1} &= 0.50 \, \rho_{\text{SEP}}^{(2)} + (1-0.50) \, \rho_{\text{PPTES}}^{(2)} \\
\sigma_{2} &= 0.55 \, \rho_{\text{SEP}}^{(2)} + (1-0.55) \, \rho_{\text{PPTES}}^{(2)} \\
\sigma_{3} &= 0.58 \, \rho_{\text{SEP}}^{(2)} + (1-0.58) \, \rho_{\text{PPTES}^{(2)}}
\end{split}
\end{equation}
where  $\rho_{SEP}, \rho_{PPTES}, \sigma_0, \sigma_1, \sigma_2$ and $\sigma_3$ are as follows,
\[
\begin{aligned}
\rho_{SEP}^{(2)} &= \frac{1}{8}\Big(
|00\rangle\langle00| + |00\rangle\langle33| + |33\rangle\langle00| + |33\rangle\langle33|\\[4pt]
&\qquad\; + |03\rangle\langle03| + |03\rangle\langle30| + |30\rangle\langle03| + |30\rangle\langle30|\\[4pt]
&\qquad\; + |11\rangle\langle11| + |11\rangle\langle22| + |22\rangle\langle11| + |22\rangle\langle22|\\[4pt]
&\qquad\; + |12\rangle\langle12| + |12\rangle\langle21| + |21\rangle\langle12| + |21\rangle\langle21|
\Big).
\end{aligned}
\]
\[
\begin{aligned}
\rho_{PPTES}^{(2)} &= \frac{1}{4}\big(
|00\rangle\langle00| + |00\rangle\langle11| + |00\rangle\langle22| + |00\rangle\langle33|\\[4pt]
&\quad\; + |11\rangle\langle00| + |11\rangle\langle11| + |11\rangle\langle22| + |11\rangle\langle33|\\[4pt]
&\quad\; + |22\rangle\langle00| + |22\rangle\langle11| + |22\rangle\langle22| + |22\rangle\langle33|\\[4pt]
&\quad\; + |33\rangle\langle00| + |33\rangle\langle11| + |33\rangle\langle22| + |33\rangle\langle33|
\big).
\end{aligned}
\]
By using the theorem (\ref{thm: condition_for_PPT_for_conv_Comb_rho}), it can be easily shown that the states $\sigma_{i}$'s $(i=0,1,2,3)$ are PPT entangled states.
In this case, the value of $1-Q$ comes out to be 1.0007, which is greater than $0$, and therefore we can conclude that the state $\rho_{c}$ defined with the tensor product of four maximally entangled states and four PPT entangled states, is useful in quantum cryptography.\\
\end{example}
\section{Conclusion}
\label{sec-7:Conclusion}
 To summarize, we explored new approaches to detect PPT states by examining the second-order moment ($p_2$) of the partial transpose, and their relation to the system’s dimension. We derived a condition that helps to identify PPT states and noted its corollary for the sufficient condition for entangled states. Additionally,
 we provided a lower bound on $p_2$, complementing the existing relation $p_2^2 \leq p_3$, to offer further insight into the properties of the second-order moment of the partial transpose of the PPT quantum states. We also studied convex combinations of separable and PPT entangled states, detectable by a witness operator $W$, and derived a condition on the mixing parameter $p$ that supports bound entanglement. Lastly, we introduced a class of states formed by sums of tensor products of Bell states and PPT entangled states, which show potential for quantum cryptography due to their positive key rates. These
 results contribute to the study of entanglement and its applications, and future work could investigate their extensions to multipartite systems or their practical
 implementation in cryptographic settings.
\section{Data Availability Statement}
 Data sharing is not applicable to this article as no datasets were generated or analysed during the current study.

\end{document}